\documentclass[aps,epsfig,twocolumn]{revtex4}
\usepackage{amsmath}
\usepackage{epsfig}
\begin{document}
\title{An effective theory of Feshbach resonances and many-body properties of Fermi gases}
\author{G.\ M.\ Bruun$^1$ and  C.\ J.\ Pethick$^2$\\  
$^1$Niels Bohr Institute, Blegdamsvej 17, DK-2100 Copenhagen \O,
Denmark\\
$^2$Nordita, Blegdamsvej 17, DK-2100 Copenhagen \O, Denmark}
\begin{abstract} 
For calculating low-energy properties of a dilute gas of atoms  
interacting via a Feshbach resonance, we develop an effective theory in which
the parameters that enter are an atom-molecule coupling strength and the
magnetic moment of the molecular resonance. We demonstrate that for resonances
in the fermionic systems $^{6}$Li and $^{40}$K that are under experimental investigation, the
coupling is so strong that many-body effects are appreciable even when the
resonance lies at an energy large compared with the Fermi energy.  We 
calculate a number of many-body effects, including the effective
mass and the lifetime of atomic quasiparticles in the gas.  
\end{abstract}
\maketitle

Due to the enormous progress made in trapping and cooling atomic gases, it  is now
possible to create degenerate Fermi gases \cite{Experiments,Loftus}.
The fact that interactions between atoms can be varied essentially at will by
exploiting Feshbach resonances makes these systems candidates for exploring experimentally strongly
correlated Fermi systems, in which the magnitude of the scattering length $a$ is larger than the inter-particle
separation. Many aspects of such systems have been considered, 
among them the equation of state \cite{heiselberg} and the cross-over between the BCS
state and a Bose-Einstein condensate~\cite{holland, griffin}. 

In this paper we begin by showing how, in the spirit of Landau Fermi-liquid 
theory and effective field theories, low energy Feshbach resonances 
may be characterized by a small number of phenomenological parameters. Related
 ideas have been proposed in Refs.\ \cite{holland,duine}, and here we provide a compact 
quantitative formulation. To do this we give a field-theoretic
account of a Feshbach resonance in a two-atom system which may readily be
generalized to the many-body problem. Next we examine the
conditions under which the Feshbach resonance induces strong correlations in a 
gas of fermionic atoms.  Finally, we calculate a number of many-body
properties.

The first step in formulating an effective low-energy theory is to identify the
relevant degrees of freedom. These we shall take to be two species ($\sigma=\uparrow\downarrow$) 
of atoms both of mass $m$, and a single low-lying molecular state
 whose energy relative to that of two stationary atoms in the open channel will be denoted by $2\nu$.
It is important to remark that this molecular state consists partly of a ``bare'' molecule in a closed
 channel (or channels), together with a 
dressing cloud of high-energy atom pairs in the open channel. Dressing of the molecular state by low-momentum pairs, which gives rise to threshold behavior, 
will be calculated explicitly.  In addition, the matrix element $g$ describing coupling of 
 atoms to the molecular state is different from the bare coupling. As an example, we calculate the width parameter 
$\Delta B$ which enters the standard expression $a(B)=a_{\rm bg}[1-\Delta B/(B-B_0)]$ for the scattering length
 $a$ as a function of magnetic field $B$ near a 
Feshbach resonance. Here $a_{\rm bg}$ is the background
 scattering length due to non-resonant processes.
 The resonant contribution to the effective atom-atom interaction at zero energy
is $-g^2/(2\nu)$.  This in turn is equal to $4\pi \hbar^2 a_{\rm res}/m$, where $a_{\rm res}$ is 
the resonant contribution to the scattering length, from which it follows that   
\begin{equation}
a_{\rm bg}\Delta B=\frac{m}{4\pi \hbar^2}\frac{g^2}{\Delta \mu}.
\label{gvac}
\end{equation}
Here 
$\Delta \mu=2\partial\nu/\partial B $ is the difference in magnetic moments between two atoms and the 
molecule.  The result (\ref{gvac}) is independent of the detailed microscopic model, provided there 
is only one low-lying molecular state.

We now demonstrate how the parameters $2\nu$ and $g$ entering the low-energy theory
are related to bare quantities  for an explicit 
microscopic 
model, with the Hamiltonian 
\begin{gather}
\hat{H}=\sum_{{\mathbf{k}},\sigma}\epsilon_k\hat{a}^\dagger_{{\mathbf{k}}\sigma}\hat{a}_{{\mathbf{k}}\sigma}+
\sum_{{\mathbf{p}},{\mathbf{k}}}\frac{g_{\rm bare}(k)}{\sqrt{\mathcal{V}}}
[\hat{b}^\dagger_{\mathbf{p}}\hat{a}_{{\mathbf{p}}/2+{\mathbf{k}}\uparrow}
\hat{a}_{{\mathbf{p}}/2-{\mathbf{k}}\downarrow}+ {\rm h.c}]
\nonumber\\
+\sum_{\mathbf{p}}E_p^{\rm bare}\hat{b}^\dagger_{\mathbf{p}}\hat{b}_{\mathbf{p}}
+\sum_{{\mathbf{k}},{\mathbf{q}},{\mathbf{q}}'}\frac{V({\mathbf{q}},{\mathbf{q}}')}{\mathcal{V}}
\hat{a}^\dagger_{{\mathbf{k}}+{\mathbf{q}}\uparrow}
\hat{a}_{{\mathbf{k}}-{\mathbf{q}}\downarrow}^\dagger\hat{a}_{{\mathbf{k}}-{\mathbf{q}}'\downarrow}
\hat{a}_{{\mathbf{k}}+{\mathbf{q}}'\uparrow}.
\nonumber
\end{gather} 
Here $E_p^{\rm bare}=2\nu_{\rm bare}+p^2/4m $, where $2\nu_{\rm bare}$ is the energy of a bare
molecule with momentum zero 
measured with respect to the energy of a pair of atoms at rest in the open
channel, $\epsilon_k=k^2/2m$ is the kinetic energy of an atom, 
$g_{\rm bare}(k)$ is the bare molecule-atom coupling matrix element, $V$ is the non-resonant 
interaction between atoms, and 
${\mathcal{V}}$ is the volume of the system. 
The quantity $2\nu_{\rm bare}$ cannot be measured directly because the energy of the 
molecule is affected by coupling to the open channel, so we  
eliminate it in favor of $\nu$, the energy of the dressed molecule.  
The energy of a molecule at rest in a vacuum is given by the poles of the  molecule propagator 
$D_{\rm{vac}}(\omega)=[\omega-2\nu_{\rm bare}-\Pi_{\rm{vac}}(\omega)]^{-1}$, where in
a compact matrix notation the vacuum molecule 
self energy has the form
$\Pi_{\rm{vac}}(\omega)=g_{\rm bare}^\dagger
G_{\rm{vac}}(\omega)g_{\rm{vac}}(\omega)$~\cite[Fig.\ 3]{duine} and 
$G_{\rm{vac}}(\omega)=(\omega-k^2/m)^{-1}$.
The quantity
$g_{\rm{vac}}(k,\omega)$ 
is the effective coupling between two atoms and a bare molecule in a
vacuum, including effects of $V$. It is given explicitly
  by   $g_{\rm{vac}}(k,z=k^2/m-i\delta)=\langle\phi_{\rm m}|\hat{H}|\phi_k^-\rangle$,
where
$|\phi_{\rm m}\rangle$ is the bare
molecular state and $|\phi_k^-\rangle$ the scattering state for two atoms with relative momentum $2{\mathbf{k}}$
  and energy $k^2/m$
in the center-of-mass frame interacting via $V$ alone \cite{duine}. We separate out the threshold behavior by writing 
$\Pi_{\rm{vac}}(\omega)-\Pi_{\rm{vac}}(0)=g^2_{\rm vac,0}m^{3/2}\sqrt{-\omega}/(4\pi)+\Delta\Pi_{\rm{vac}}(\omega)$, 
\begin{equation}
\Delta\Pi_{\rm{vac}}(\omega)=m^2\omega\int_0^\infty\frac{dk}{2\pi^2}\frac{g_{\rm{vac}}(k,\omega)g_{\rm{vac}}(k,0)
-g_{\rm vac, 0}^2}{m\omega-k^2},\nonumber
\end{equation}
and $g_{\rm vac, 0}\equiv g_{\rm vac}(0,0)$. 
The  $\sqrt{-\omega}$ term is due to the threshold in the density of states
for the open channel 
at $\omega=0$. For positive energy it
 gives rise to damping of the molecular resonance, while for $\omega <0$
it results in the binding 
energy of the molecular state varying as $1/a^2$ close to the resonance. 
We shall assume that 
frequencies of interest are small compared with the characteristic
frequencies that enter the coupling matrix elements 
$g_{\rm{vac}}(k,\omega)$. One then finds
$D_{\rm vac}^{-1}(\omega)=\omega/\tilde{z} - 2\nu_{\rm bare}
-\Pi_{\rm{vac}}(0)-{g^2_{\rm vac, 0}}m^{3/2}\sqrt{-\omega}(4\pi)^{-1}$, where 
$\tilde{z}^{-1}=1-\partial \Delta\Pi_{\rm{vac}}(0)/\partial 
\omega$. Physically, $\tilde{z}$ is the renormalization factor for the molecular resonance when low-energy 
atom pairs in the open channel are neglected. 
 This shows that the resonance energy, which is defined by ${\rm Re}[D_{\rm vac}^{-1}(2\nu)]=0$, is given by 
\begin{equation}
2\nu=\tilde{z}[2\nu_{\rm bare} +\Pi_{\rm{vac}}(0)]
\label{molenergy}
\end{equation}
for $2\nu_{\rm bare}+\Pi_{\rm{vac}}(0)>0$.
Thus, the energy of the resonance is shifted, and the difference between the magnetic
moments of the two atoms in the open channel relative to that of the molecule 
is $\Delta\mu=2\partial \nu/\partial B=\tilde{z}\Delta\mu_{\rm bare}$, 
$\Delta \mu_{\rm bare}=2\partial\nu_{\rm bare}/\partial B $ being the difference in magnetic
moments between two atoms in the open channel and the 
{\it bare} molecular state.
The energy of the molecular state is given for small
$B-B_0$ by $2\nu=\Delta\mu(B-B_0)$. 
For negative $\omega$ one finds  $1-\partial \Pi_{\rm{vac}}(\omega)/\partial 
\omega =\tilde{z}^{-1} +{g^2_{\rm vac, 0}}m^{3/2}/(8\pi |\omega|^{1/2})$.  The
singularity
 for $\omega\rightarrow 0^-$ reflects the fact that the 
wave function of a very weakly bound molecular state consists mainly of low-energy
pairs of atoms in the open channel, 
with only a very small closed-channel component~\cite{Braaten}.

According to standard theory, the resonant contribution $a_{\rm res}$ to the scattering length is 
$4\pi a_{\rm res}/m={g^2_{\rm vac, 0}} D_{\rm vac}(\omega=0) = -{g^2_{\rm vac, 0}}/[2\nu_{\rm bare}+\Pi_{\rm vac}(0)]$ \cite{duine}. 
With the help of 
Eq.\ (\ref{molenergy}), this may be  written as  $-\tilde{z}{g^2_{\rm vac, 0}}/(2\nu)$, in agreement with Eq.\ (\ref{gvac})
if the effective coupling between  
atoms and a dressed molecule is given by 
\begin{equation}
g=\sqrt{\tilde{z}}g_{\rm vac, 0}.
\end{equation}
The discussion so far applies equally well for bosons and fermions, apart
from additional statistical factors if the two atomic species are 
identical.  

As an application, we consider in the rest of the paper a
two-component Fermi gas. 
We assume that the populations of the atomic species
are equal and denote the total density by $n$. 
A dimensionless measure of the importance  of the many-body effects due to the resonant 
interaction  is $k_{\rm F}|a_{\rm res}|$,
where $k_{\rm F}$ is the Fermi momentum,  and we now estimate this quantity for systems of experimental 
interest. Feshbach resonances for fermionic atoms have been
reported for $^{40}$K and  $^6$Li. 
For $^{40}$K, there is a Feshbach resonance for atoms in the states $|9/2,-9/2\rangle$ and $|9/2,-7/2\rangle$
with parameters $B_0\simeq 201.5$ G, $\Delta B\approx 8$ G, and
$a_{\rm bg}=164a_0$~\cite{Bohn,Loftus}. The only other state in the ground-state manifold to which the open 
channel can couple by the central part of the interaction is $|9/2,-9/2\rangle$ and $|7/2,-7/2\rangle$.
Diagonalization of the hyperfine Hamiltonian yields $\Delta\mu_{\rm bare}=1.78\mu_{\rm B}$
for $B\approx 200$ G whereas a coupled-channel
calculation gives $\Delta\mu=0.118 \mu_{\rm B}$ for the molecule dressed only 
by high-energy pairs~\cite{Bohnprivate}. From this, we
obtain $\Delta\mu/\Delta\mu_{\rm bare}=0.067$ for $^{40}$K, which demonstrates that the
renormalization of the molecule energy due to the coupling to the open
channels is significant even away from threshold. With these parameters we find
 $k_{\rm F}|a_{\rm res}|\simeq 15n_{12}^{-1/3}\epsilon_{\rm F}/\nu$
where $n_{12}$ is  the density in units of $10^{12}$ cm$^{-3}$ and $\epsilon_{\rm F}=k_{\rm F}^2/2m$.
The dimensionless coupling is therefore quite strong  for $^{40}$K.
For $^6$Li we do not know the value of $\Delta\mu$, so as a first
approximation we take
$\Delta\mu_{\rm bare}\sim2\mu_{\rm B}$~\cite{holland,Houbiers} appropriate for the bare molecule.
Using the parameters $\Delta B\approx 185$ G and  $a_{\rm bg}\approx2160a_0$, 
we find $k_{\rm F}|a_{\rm res}| = 1.1\times10^4 n_{12}^{-1/3}  \epsilon_{\rm F}/\nu_{\rm bare}$. 
At a frequency $\omega$, the resonant contribution to the 
effective interaction between atoms in the open channel is given by $g^2/(\omega -2\nu)$. 
Since in calculations of many-body effects
typical energies are of order  $\epsilon_{\rm F}$, this interaction may be treated as being independent of frequency provided 
$2\nu \gg \epsilon_{\rm F}$. Our  discussion shows that even when this condition is satisfied, 
systems can be strongly coupled with  $k_{\rm F}|a_{\rm res}| \gg 1$. 

The next question we address is how the molecule properties are modified by the presence
of other atoms. Such effects have recently been considered by Combescot \cite{combescot}. 
The molecule propagator in a medium is given by 
$D(p,\omega)^{-1}=\omega-E_p^{\rm bare}+2\mu-\Pi(p,\omega)\equiv
D_0(p,\omega)^{-1}-\Pi(p,\omega)$
 where $\Pi(p,\omega)$ is the many-body self energy and $2\mu$ is the
chemical potential of a molecule. As a simple approximation 
for $\Pi$ we take the same processes as in the calculation of $\Pi_{\rm
vac}$ except that the 
atom propagator is that in the presence of a medium, rather than that in a
vacuum 
\cite{pandharipande}. This yields $\Pi(p,\omega)=g_{\rm bare}^\dagger G_{\rm mb}^{(2)}(p,\omega)g_{\rm mb}(p,\omega)$
where $G^{(2)}_{\rm mb}({\mathbf{p}},{\mathbf{k}},\omega)=[n_{\rm F}(\xi_{{\mathbf{p}}/2+{\mathbf{k}}})+
n_{\rm F}(\xi_{{\mathbf{p}}/2-{\mathbf{k}}})-1]/(\xi_{{\mathbf{p}}/2+{\mathbf{k}}}+
\xi_{{\mathbf{p}}/2-{\mathbf{k}}}-\omega-i\delta)$,
with $\xi_k=\epsilon_k-\mu$,
describes the propagation of a pair of atoms with total momentum
${\mathbf{p}}$, energy $\omega$, and 
relative momentum $2{\mathbf{k}}$ in the presence of the other atoms. Here $n_{\rm F}(x)=[\exp(\beta x)+1]^{-1}$
with $\beta^{-1}=k_{\rm B}T$.
 The  many-body atom-molecule interaction in the ladder approximation is given by 
$g_{\rm mb}(p,\omega)=g_{\rm{vac}}(\tilde{\omega})[1-\Delta G_{\rm mb}T_{\rm{bg}}(\tilde{\omega})]^{-1}$ with 
 $\Delta G_{\rm mb}=G^{(2)}_{\rm mb}(p,\omega)-G_{\rm{vac}}(\tilde{\omega})$
and $\tilde{\omega}=\omega+2\mu-p^2/4m$.
$T_{\rm{bg}}$ is the vacuum $T$-matrix for the background interaction $V$ 
which for low energies is given by $T_{\rm{vac}}=4\pi a_{\rm bg}/m$. One
therefore  has 
$g_{\rm mb}(p,\omega)=g_{\rm{vac}}(\tilde{\omega})+{\mathcal{O}}(k_{\rm F}a_{\rm bg})$.
For clarity,  we ignore  such correction 
terms in the following discussion of the molecule properties assuming $k_{\rm F}|a_{\rm bg}|\ll 1$ and thus
$g_{\rm mb}(p,\omega)=g_{\rm{vac}}(\tilde{\omega})$. 
 Expressing $\Pi(p,\omega)$ in terms of $\Pi_{\rm vac}(p,\omega)$ we obtain
\begin{equation}\label{D}
\tilde{z}D^{-1}(p,\omega)=\omega-E_p-g^2\frac{m^{3/2}}{4\pi}\Theta(-\tilde{\omega})\sqrt{|\tilde{\omega}|}-\tilde{\Pi}(p,\omega).
\end{equation}
Here $E_p=p^2/4m+2\nu-2\mu$ is the difference between the molecule energy and the chemical potential
when dressing by low-energy atoms is neglected. The term 
\begin{equation}\label{pirenorm}
\frac{\tilde{\Pi}(p,\omega)}{g^2}=\int
\frac{d^3k}{(2\pi)^3}\left[G^{(2)}_{\rm mb}({\mathbf{p}},{\mathbf{k}},\omega)-G_{\rm vac}(k,\tilde{\omega})
\right]
\end{equation}
gives the contributions to the self energy coming from the presence of the
atoms~\cite{Note}. 
In calculating $\tilde{\Pi}$ we have replaced  $\sqrt{\tilde{z}}g_{\rm vac}(k,\omega)$ by $g=\sqrt{\tilde{z}}g_{\rm vac,0}$
since contributions from high momentum states, where the difference between $g_{\rm vac}(k, \omega)$ and $g_{\rm vac,0}$
 is appreciable, are cut off by the Fermi functions. This approximation
will be used in the rest of the paper.
The importance of Eqs.\ (\ref{D}) and (\ref{pirenorm}) is that
they show explicitly how molecular properties 
at low energies may be expressed solely in terms of experimentally
measurable quantities $g$ and $\Delta \mu$, 
without arbitrary assumptions about the behavior of the bare coupling matrix
elements for high momenta.

We turn now to properties of atoms described by the propagator 
$G(p,\omega)^{-1}=G_0(p,\omega)^{-1}-\Sigma(p,\omega)$ 
with $G_0(p,\omega)^{-1}=\omega-\xi_p$. The atom self energy is given in the 
ladder approximation  by 
$\Sigma(p,i\omega_n)=\beta^{-1}{\rm Tr}[\Gamma({\mathbf{K}},{\mathbf{q}},
{\mathbf{q}},i\omega_n+i\omega_m)G_0(k,i\omega_m)]$ 
where ${\mathbf{K}}={\mathbf{p}}+{\mathbf{k}}$, ${\mathbf{q}}=({\mathbf{p}}-{\mathbf{k}})/2$ and the 
trace denotes a sum over $\omega_m=(2m+1)\pi k_{\rm B}T$ and integration over ${\mathbf{k}}$ \cite{fetter}.
The matrix $\Gamma(p,\omega,k,k')$ obeys the equation 
$\Gamma(p,\omega)=V_{\rm eff}(p,\omega)+V_{\rm eff}(p,\omega)
G^{(2)}_{\rm mb}(p,\omega)\Gamma(p,\omega)$
with $V_{\rm eff}(p,\omega)=V+g_{\rm bare}^\dagger D_0(p,\omega)g_{\rm bare}$. 
After some manipulations $\Sigma$ can be expressed as 
\begin{gather}
\Sigma(p,\omega)=\Sigma_{\rm bg}(p,\omega)+\int\frac{d^3k}{(2\pi)^3}g_{\rm
mb}({\mathbf{k}}+{\mathbf{p}}, \frac{{\mathbf{k}}-{\mathbf{p}}}{2},\omega)^2\nonumber\\
\times\int\frac{d\epsilon}{2\pi}[n_{\rm F}(\epsilon)A_{\rm a}(k,\epsilon)D({\mathbf{k}}+{\mathbf{p}},\epsilon+\omega)\nonumber\\
-n_{\rm B}(\epsilon+\omega)G(k,\epsilon)A_{\rm m}({\mathbf{k}}+{\mathbf{p}},\epsilon+\omega)]\label{SelfFermi}
\end{gather}
with $n_{\rm B}(x)=[\exp(\beta x)-1]^{-1}$ 
and $A_{\rm m}(k,\omega)=-2{\rm Im}D(k,\omega)$ and $A_{\rm a}(k,\omega)=-2{\rm Im}G(k,\omega)$
being the molecule and atom spectral functions respectively. 
Using the low energy approximations $T_{\rm{vac}}=4\pi
a_{\rm bg}/m$ and $g_{\rm{vac}}(k,\omega)=g_{\rm vac, 0}$, we obtain 
\begin{equation}\label{gmb}
\frac{g_{\rm mb}(\mathbf{p},\mathbf{k},\omega)}{g_{\rm vac, 0}}=1+\int\frac{d^3q}{(2\pi)^3}
\frac{\Delta G_{\rm mb}({\mathbf{p},{\mathbf{q}}},\omega)T_{\rm bg}}{1-\Delta G_{\rm mb}({\mathbf{p}},{\mathbf{q}},\omega)T_{\rm bg}}.
\end{equation}
For $T\rightarrow0$, the self energy due to the background scattering $V$ is 
given in the ladder approximation by 
$\Sigma_{\rm bg}(p,\omega)=[2k_{\rm F}^2/(3m\pi)][k_{\rm F}a_{\rm bg}+{\mathcal{O}}(k_{\rm F}a_{\rm bg})^2]$~\cite{galitskii}.
Since $D$ is given by Eqs.\  (\ref{D}) and (\ref{pirenorm}), one sees from Eqs.\ (\ref{SelfFermi})
 and (\ref{gmb}) that the atom 
self energy may be expressed in terms of $g$ and $\nu$ (or equivalently $\Delta \mu$ and $B_0$).

From $\Sigma$ one can calculate the properties of the
atom-like excitations given by the poles of $G$.                             
We begin by describing results for weak coupling, i.e. $k_{\rm F}|a_{\rm res}|\ll 1$.
  For $2\nu\gg \epsilon_{\rm F}$ and $T=0$, 
one finds that the effective mass of an excitation at the Fermi surface is given by 
\begin{equation}\label{effectivemass}
\frac{m^*}{m}=1+\frac{g^2}{16\nu^2}n+(21\ln 2-3)
\frac{(T_{\rm vac}-g^2/2\nu)^2 n^2}{40\epsilon_{\rm F}^2}
\end{equation}
and the wave function renormalization factor $Z$ by
\begin{equation}\label{renormconstant}
Z^{-1}(p)=1+\frac{g^2}{8\nu^2}n+\frac{9\ln 2}{8}\frac{(T_{\rm vac}-g^2/2\nu)^2n^2}{\epsilon_{\rm F}^2}.
\end{equation}
The second terms in Eqs.\ (\ref{effectivemass})-(\ref{renormconstant}) come from the 
frequency dependence of the molecule propagator, 
while the final ones are identical with Galitskii's results for a dilute
Fermi gas with short-range interactions parametrized by a scattering length 
$a=a_{\rm bg}+a_{\rm res}$~\cite{galitskii}.
If one neglects $a_{\rm bg}$, the last two terms in the expression for $m^*$
 have the same dependence on $\nu$, and the last term is 
$\approx 2k_{\rm F}|a_{\rm res}|\nu/\epsilon_{\rm F}$ times
the second one.
Thus for small $g$, the second term dominates over the last one while, for
the resonances in $^6$Li and $^{40}$K considered 
above, the last term is the more important at densities of experimental
interest. For example, for the $^{40}$K resonance, the ratio 
is $\sim 30 n_{12}^{-1/3}$, which demonstrates that the effective interaction 
is equivalent to a contact interaction between atoms 
of strength $g^2/2\nu$, provided that $\nu\gg \epsilon_{\rm F}$. 
  Note that for the resonances in $^6$Li and $^{40}$K
considered in this paper, one can have 
$k_{\rm F} |a|\gtrsim 1$ even when $\nu\gg \epsilon_{\rm F}$. Thus, the gas can be
strongly interacting due to the Feshbach resonance while
the interaction still can be regarded as effectively instantaneous with a scattering length $a$,
in agreement with the earlier order of magnitude estimates.  

The imaginary part of the atom self energy may be calculated in a similar
manner. There is a  contribution of 
order $g^2$ coming from creation of real molecules.  This is given for $T=0$ by
${\rm Im}[\Sigma(p,\xi_p)]=8k_{\rm F}a_{\rm res}\nu^{3/2}\epsilon_{\rm F}^{-1/2}\times$
$\{1-p/\sqrt{128m\nu}+[\epsilon_{\rm F}/(4\nu)-1](2m\nu)^{1/2}/p\}$
for $\sqrt{8m\nu}-\sqrt{2m\epsilon_{\rm F}}\le p\le\sqrt{8m\nu}+\sqrt{2m\epsilon_{\rm F}}$ and zero otherwise.
For $2\nu\gg \epsilon_{\rm F}$, this on-shell process is unimportant for
atoms with $p\lesssim{\mathcal{O}}(k_{\rm F})$. 
For momenta outside the range for which the above process is allowed,
damping is due to
creation of particle-hole pairs. One finds for $|p-k_{\rm F}|/k_{\rm F}\ll 1$ and $T\ll T_{\rm F}$
to lowest non-trivial order in $a_{\rm bg}$ and $g$ that 
\begin{equation}
\frac{{\rm Im}\Sigma(p,\xi_p)}{\epsilon_{\rm F}}=-\frac{2}{\pi}(k_{\rm F}
a)^2\left[(1-\frac{p}{k_{\rm F}})^2+(\frac{\pi k_{\rm B}
T}{\epsilon_{\rm F}})^2\right]
\label{imsigmaweak}
\end{equation}  
where again $a=a_{\rm bg}+a_{\rm res}$.
For $T=0$, this result agrees with the expression derived  by
Galitskii 
for a gas of particles interacting via a short range interaction with
scattering length $a$~\cite{galitskii}.

We turn now to stronger couplings, and we shall neglect $a_{\rm bg}$ compared
with $a_{\rm res}$.  
The important ingredient in  Eq.\ (\ref{SelfFermi}) is the molecule spectral function 
\begin{equation}\label{ImDgeneral}
A_{\rm m}(p,\omega)=\frac{-2{\rm Im}\tilde{\Pi}(p,\omega)}{[\omega-E_p-{\rm Re}\tilde{\Pi}(p,\omega)]^2+{\rm Im}\tilde{\Pi}(p,\omega)^2}.
\end{equation}
For calculating properties of the system at $T\lesssim T_{\rm F}$, typical energies of 
interest are of order $\epsilon_{\rm F}$. 
\begin{figure}
\centering
\epsfig{file=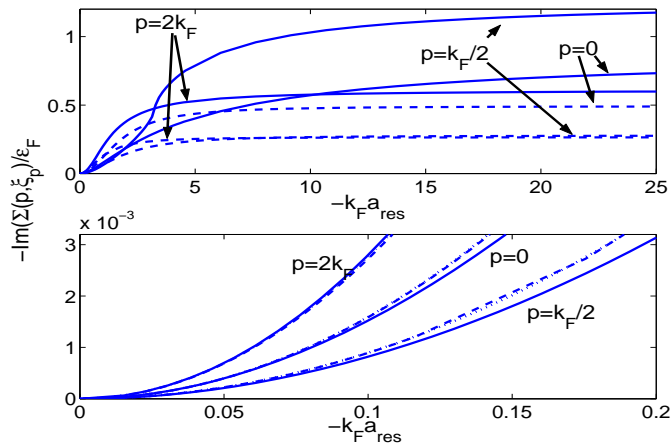,height=0.25\textheight,width=0.5\textwidth,angle=0}
\caption{$-{\rm Im}\Sigma(p,\xi_p)/\epsilon_{\rm F}$ as a function of $k_{\rm F}|a_{\rm res}|$ for various momenta. 
The solid curves are numerical results based on 
Eqs.\ (\ref{SelfFermi}) and (\ref{ImDgeneral}), the dashed curves are the results if the effect of the medium on the scattering 
amplitude is neglected, and the dotted curves in Fig.(b) are the Galitskii results [Eq.(\ref{imsigmaweak})]. 
All these results become identical for $k_{\rm F}|a_{\rm res}|\ll 1$.}
\label{Unitarity}
\end{figure}
 For such energies, ${\rm Im}\, \tilde\Pi$ is of order $ g^2m k_{\rm F}$, and therefore for large $\nu$ ($k_{\rm
F}|a_{\rm res}| \ll 1$), the 
denominator in Eq.\ (\ref{ImDgeneral}) may be replaced by $(2\nu)^2$, and one recovers the
weak-coupling result (\ref{imsigmaweak}).  With
 decreasing molecule energy $\nu$, the imaginary term becomes increasingly
important. In Fig.\ \ref{Unitarity}  we show numerical 
results for the quasiparticle damping rate $-{\rm Im} \Sigma(p, \xi_p)$ based on Eqs.\ (\ref{SelfFermi}) and (\ref{ImDgeneral}),
for $T=0$ and a coupling strength $g$ 
appropriate for the $^{40}$K resonance discussed above.  For $k_{\rm F}|a_{\rm res}|\gg 1$,
the damping rate saturates because
the $\tilde \Pi$ terms in the denominator dominate.  For comparison, we also show the results obtained by
neglecting medium effects in the denominator. In a Boltzmann equation approach, this amounts to 
assuming that the differential cross section for two-atom scattering with relative momentum $k$ 
is given by its vacuum expression
 $a_{\rm res}^2/(1+k^2a_{\rm res}^2)$, where effective range contributions have been neglected. 
As Fig.\ \ref{Unitarity} (a)  shows, for large $k_{\rm F}|a_{\rm res}|$  the effect of the
 medium on the scattering amplitude \emph{increases} the damping rate significantly as compared to the Boltzmann 
results using the vacuum scattering rate. Remarkably, the scattering rates are substantially greater than one  
would predict if scattering cross sections were given by the unitarity limit for two particles in a vacuum.
This is due primarily 
to the reduction by Pauli blocking of the magnitude of ${\rm Im}\, \tilde \Pi$ , and hence also of the magnitude of the 
scattering amplitude.
Figure Fig.\ \ref{Unitarity} (b) also shows
that the results agree with those of Galitskii for $k_{\rm F}|a_{\rm res}|\ll 1$.

The main results of this paper are, first, that we have shown how to formulate a theory of Feshbach resonances 
that involves only low-energy observables. One of these quantities is the magnetic moment of the dressed 
molecule, which is very different from that of the bare molecule in cases of experimental interest. We have
 also shown that for Fermi gases of experimental interest,
many-body effects are strong even when the Feshbach resonance lies at energies well above the Fermi energy,
 and that for many purposes the interaction induced by the resonance may be regarded as instantaneous.
 As applications of the theory, we calculated a number of many-body properties of a two-component Fermi gas. 
One important result is that, for strong interactions, scattering rates can exceed substantially predictions based on 
unitarity-limited two-body cross sections.  

We are very grateful to John Bohn for providing us with his result for $\Delta \mu$ for $^{40}$K.


\begin{thebibliography}{99}
\bibitem{Note} $\tilde{\Pi}$ includes all contributions to the imaginary
part as we have only subtracted the 
real part of $\Pi_{\rm{vac}}$.
\end{thebibliography}
\end{document}